# Unresolved Wind-Driven Shells and the Supersonic Velocity Dispersion in Giant HII Regions

Guillermo Tenorio-Tagle[1],

Casiana Muñoz-Tuñón[1] and

Roberto Cid-Fernandes[2,3]

1 Instituto de Astrofísica de Canarias, E-38200 La Laguna, Tenerife. Canary Islands, Spain

2 Institute of Astronomy, Madingley Road, Cambridge CB3 OHA UK

3 Royal Greenwich Observatory, Madingley Road, Cambridge CB3 OEZ, UK

# ABSTRACT

The presence of giant shells or loops in giant HII regions are clear witness of the mechanical energy input from massive stars. Here we evaluate the impact that winds may have on the structure of giant nebulae and on their supersonic velocity dispersion. We follow the suggestion from Chu & Kennicutt (1994) to see if a combination of a large number of unresolved wind-driven shells caused by massive stars could produce the integrated broad Gaussian profiles typical of giant HII regions.

The results, accounting for a wide range of energies, densities and velocity or age of the expanding shells, show that supersonic Gaussian profiles may arise only from a collection of unresolved wind-driven shells if the shells present a peculiar velocity distribution which implies a strongly peaked age distribution leading to an awkward star formation history. On the other hand, a uniform distribution of ages originates profiles with a flat-topped core defined by the terminal shell velocity and a steep decay as $v^{-6}$ up to the largest detectable shell speed. Thus, supersonic profiles can arise only if the final speed of the unresolved shells is supersonic. This implies an equally supersonic random speed of motions in the ionized gas disrupting the shells before they slow down to subsonic speeds. It also implies a mechanism, independent of the shells caused by massive stars, responsible for the supersonic stirring of the background medium. These facts together with the conditions for shells to remain unresolved to present day devices (energies, final speeds and ages), indicate that the winds may be produced by low-mass stars. In this latter case, if the sources move supersonically in the gravitational potential of the whole system they could stir the gas, with their cometary bow shocks, to a $\sigma_{gas} \sim \sigma_{stars}$ causing a supersonic local random speed of motions within the system.

## 1. Introduction

One of the most debated issues in the field of giant HII regions is without doubt the nature of their supersonic gas motions. It is well understood that an energy source is necessary to replenish, at all times, the dissipation caused by radiative cooling behind the multitude of shock waves that should result from the supersonic motions. In this respect the integrated mechanical energy deposited by the large number (hundreds or thousands) of massive stars, in most cases inferred from photoionization requirements, has been presumed to be the main mechanism driving supersonic turbulence (Dyson 1979; Rosa & Solf 1984; Hippelein & Fried 1984; Chu & Kennicutt 1994, hereafter CK). Given the apparent association between molecular gas and the giant nebulae, champagne flows have also been thought to be responsible of the supersonic random motions (Skillman & Balick 1984; Hunter & Gallagher 1985).

On the other hand, the supersonic velocity dispersion ($\sigma$) in giant HII regions has been shown to correlate with the size ($\sigma \sim \text{size}^{0.5}$) and luminosity ($\sigma \sim L^{0.25}$) of the giant nebulae (Terlevich & Melnick 1981), and the scatter in the data is strongly reduced if one considers instead of the core or the halo dimensions, or a combination of the two, the "kinematic radius" (Muñoz-Tuñón 1994), *i.e.* the dimension of the area where high-resolution studies warrant a consistently uniform supersonic Gaussian profile independent of location. These facts have no apparent explanation within the framework of the energetics caused by massive stars. Terlevich & Melnick (1981) noted that the same correlations apply to virialized systems such as globular clusters and the nuclei of elliptical galaxies, and from this they suggested that gravitational energy must be the main source that drives supersonic gas motions, accounting in this way for the observed correlations. In their original suggestion Terlevich & Melnick proposed as a possibility a large number of ionized condensations moving supersonically in the virialized (gas + stars) system, and called it the "gravitational model". The model however requires a mechanism to replenish the ionized fragments, given that champagne effects would disperse condensations in a time much shorter than the HII region life-time. Their observational result, however, has been thoroughly confirmed and also extended to HII galaxies (see Melnick 1992 and references therein).

Turbulence has also been suggested as a mechanism to explain both the supersonic width and the observed relationship between velocity dispersion and giant HII region size (Roy *et al.* 1986). There are however, several important issues regarding this suggestion. For instance, it is not clear from the velocity structure that the observed random motions are evidence of turbulence, and, as observed by Castañeda (1988), one is not even sure that this "turbulence" is the same as that described by the conventional theories of fluid mechanics (see also Melnick *et al.* 1987).

A more recent explanation of the giant HII region empirical correlations was derived as a direct consequence of the formation of massive stellar clusters ( Tenorio-Tagle , Muñoz-Tuñón & Cox 1993). There the supersonic velocity dispersion arises from the virialized motion of an ensemble of recently formed low-mass stars which undergoing low-energy winds, produces a sufficient number of cometary bow shocks to stir the surrounding matter to a $\sigma_{gas} \sim \sigma_{stars}$ and thus locally enhance the random speed of motions to a supersonic value. The virialized passage of the cometary shocks is responsible for the stabilization of the collapse and also ultimately for the observed $\sigma$ *vs* size and luminosity correlations found by Terlevich & Melnick (1981). In the case of massive clusters ($M > 10^5$ $M_\odot$), the motion of the virialized stars and of the random speed of motions the parent cloud remain supersonic ($> 10\,\mathrm{km\,s^{-1}}$) even after massive star formation and photoionization. In the cometary stirring model, during this phase the parent cloud, formerly only subjected to the cometary passage of the many low-mass wind-driven sources becomes disrupted by the mechanical energy of the massive members of the cluster. And thus, as the volume affected by the mechanical energy from massive stars grows with time, matter is removed from the location where the cometary sources could stir it, finally losing the information about the true origin of the supersonic velocity dispersion.

Recently, new data sets on 30 Doradus (CK) and NGC 604 (Sabalisck *et al.* 1995) have refreshed the interest on the topic. The NGC 604 2D Fabry-Perot data set spatial sampling is superior to any data set ever taken of 30 Doradus, including that of Smith & Weedman (1970) or that of CK, and it surely is useful to clarify the nature of the giant HII region supersonic gas motions. In the case of 30 Doradus the observations of CK revealed a number of major kinematic subsystems (shells, rings, filaments, networks of shells, etc.) and a line splitting of 20-150 $\mathrm{km\,s^{-1}}$. From this, the authors concluded that the most important physical mechanism which produces the global velocity dispersion is that which accelerates the shells, namely stellar winds from OB associations. The result from NGC 604 seems totally different. There, all lines clearly arising from shells and or filaments, apart from being strongly asymmetric or even split lines, are all very low intensity lines. On the other hand, lines arising from the two brightest emitting knots in NGC 604 are broad well behaved Gaussians implying a supersonic velocity dispersion intrinsic to the brightest knots, which do not seem at all dominated by the action of the many strong stellar winds operating in the region.

Here we evaluate the impact that winds may have on the structure of giant HII regions and on the supersonic velocity dispersion. Section 2 analyses the outcome of recent high resolution studies on the nearest giant HII regions. We follow the suggestion of CK looking to see whether the integrated profile caused by hundreds of unresolved wind-driven shells could combine to produce broad Gaussian profiles. Section 3 derives the emission profile produced by single and multiple shells, accounting for a wide range of powering energies, background densities and velocity or age range of the expanding structures. Section 4 lists our conclusions.

## 2. Supersonic turbulence in the nearest giant HII regions

The following discussion is based on recent studies of the two closest giant HII regions, 30 Doradus at a distance of 50 kpc in the LMC, and NGC 604 in M33 at a distance of 720 kpc. The two objects have been the subject of recent high-resolution spectroscopy and the corresponding outputs are used here to draw firm conclusions regarding the supersonic nature of the velocity dispersion in nearby giant HII regions.

In the case of 30 Doradus, the inner $9'^2$, corresponding to the central 135 pc, were covered by more than 37 echelle spectra (CK). From these, many of the wind-driven shells and networks of shells produced by the mechanical input from massive stars were clearly identified through emission-line patterns. This led the authors to conclude that "at high spatial and spectral resolution the nebular velocity field breaks up into hundreds of discrete supersonic structures with diameters ranging from less than 1 to 100 pc, expansion speeds of 20–300 $km\,s^{-1}$, and energies of $10^{47-51}$erg."

The main point in their study is however, that the sum of the 37 spectra from the core region (some of which arise from well identified shells or sections of shells) results in a well behaved Gaussian profile (their figure 11). This led to the conclusion that "the unusually high supersonic velocity of 30 Doradus when measured on large scales is due to a combination of hundreds of discrete motions combined with a smooth underlying turbulent velocity field", and by extrapolation to conclude that "the use of global velocity profiles of giant HII regions to draw inferences about the physical origin of the gas motions is a futile attempt."

There are, however, several points that seem to have been overlooked in the analysis of CK, and that we regard as fundamental to this discussion. The background or "smooth turbulent velocity field" seems totally free or uncontaminated by the presence of any of the resolved shells, and thus one should notice that : 1) the so called smooth turbulent velocity field found as a background component in the halo, and with the same properties (line width) in the core of the region, also presents a Gaussian profile; 2) the derived $\sigma$ value of this component is also supersonic ($\sigma_{background} \sim 18$ km s$^{-1}$), and 3) such a component accounts for at least half (some 5.6 $10^{51}$ erg) of the kinetic energy present in 30 Doradus. The smooth Gaussian profile of the background component is clearly seen in their figures 9a and c, which correspond to typical spectra from the halo (far away from R136) and the central core, respectively. The two lines differ by a factor of about 400 in intensity but otherwise they are identical with a supersonic $\sigma$ value of 15–20 km s$^{-1}$. These lines are the ones CK account for with a collection of unresolved wind-driven shells. Note on the other hand that, if supersonic gas motions are due to gravity, the measured $\sigma$ value of the background component implies a total gravitational mass $M_{Total} = 220(R/pc)(\sigma/kms^{-1})$ (see Terlevich & Melnick 1981) in the range 3–6 $\times 10^6$ M$_\odot$ , if one uses R = 67.5 pc (the radius of the central 9'), and a $\sigma$ in the range 15–20 km s$^{-1}$, which is consistent with the total detected mass of the system (Melnick 1992).

From the 2D Fabry-Perot study on NGC 604 (which led to 65000 spectra in H$_\alpha$ , and 65000 in [OIII] over the central 230 pc of the nebula). Sabalisck *et al.* (1995) derived the particular sectors of the nebula that cause the global velocity dispersion value of this source. It was shown that the brightest knots (knots 1 and 2) are the ones that show clearly well behaved Gaussian profiles with a $\sigma$ value ($\sim 17$ km s$^{-1}$) similar to the global velocity dispersion associated with NGC 604 ($\sigma_{604}$) by earlier single aperture studies. The two emitting knots ($\sim 7$ pc in radius) lie within the central zone of NGC 604 separated by 50 pc from each other. Many other values of $\sigma$ were found in the region, both subsonic and even larger than $\sigma_{604}$. The latter ones result from a Gaussian fit to strongly asymmetric or split lines, most likely caused by the photoionized matter swept up by the winds from the several W-R and massive stars identified in the region. However, in all cases the intensity of those lines is much smaller than that presented by the truly Gaussian profiles that emanate from either knot 1 or knot 2. Clearly, in NGC 604 the sampling of the brightest zones leads to profiles similar to the integrated global ones. This set of observations implies *the existence of a motor intrinsic to knots 1 and 2 capable of causing the same supersonic $\sigma_{604}$ value in both of them, despite their physical separation.* For the case of 30 Doradus the background gas shows the same $\sigma_{30Dor}$ both in the core and the halo implying an extended energetic mechanism in continuous operation. CK inferred by extrapolation of their integrated line profile that the supersonic Gaussian lines emanating from the background flow should also be the result of a large number (hundreds) of spatially unresolved wind-driven shells from massive stars. Here we follow their suggestion, looking to see if a large number of unresolved shells does indeed lead to the supersonic Gaussian lines typical of the background component of giant HII regions, and whether this has any further implications on the history of star formation. Our approach differs from the original one of Dyson (1979), in that here different powering energies, background densities, and age distributions of the wind-driven shells are considered.

## 3. The signature of unresolved multiple shells

From the equations of motion and evolution of wind-driven shells one can derive the profiles produced by multiple shell systems. Here it is further assumed that the predicted line profiles result from a population of wind-driven shells acting concurrently on an HII region while remaining unresolved to the high spatial resolution of present-day devices. The major goal of this exercise is to test whether such a system is capable of producing the smooth, supersonic Gaussian lines observed in giant HII regions.

## 3.1. Individual shells

The mechanical energy deposition from massive stars leads to a double shock pattern. The inner shock decelerates the wind while thermalizing its kinetic energy and the outer one sweeps up the surrounding background gas (see Weaver *et al.* 1977). The two gases are well separated by a contact discontinuity, and are kept in pressure equilibrium throughout the evolution. Thus, at the onset of strong radiative cooling, the matter swept up by the outer shock condenses into a geometrically thin outer shell. In the presence of a large ionizing photon flux, this will be kept fully ionized at a temperature $\sim \mathrm{T}_{HII}$, while contributing to an emission line spectrum typical of HII regions. Wind-driven shells around massive stars (or groups of stars) thus develop isothermal shocks, and consequently the density in the shells is $\mathcal{M}^2$ times the ISM density, where $\mathcal{M}$ is the outer-shock Mach number. Under these conditions, the equations describing the evolution of the velocity ($v_s$), outer radius ($R_s$), thickness ($\Delta R_s$), mass ($M_s$) and H$\alpha$ luminosity ($L_s$) of a single isolated shell are (*e.g.* , Dyson 1979):

$$v_s(t) = 100(\epsilon/n)^{1/5}t^{-2/5} \text{ km s}^{-1} \tag{1}$$

$$R_s(t) = 1.6(\epsilon/n)^{1/5}t^{3/5} \text{ pc} \tag{2}$$

$$\Delta R_s(t) = R_s(t)\left[1 - \left(1 - 1/\mathcal{M}(t)^2\right)^{1/3}\right] \tag{3}$$

$$M_s(t) = 0.07\epsilon^{3/5}n^{2/5}t^{9/5} \text{ M}_\odot \tag{4}$$

$$L_s(t) = 1.6 \times 10^{34}(c_{HII}/10)^{-2}nt\epsilon \text{ erg s}^{-1} \tag{5}$$

where $\epsilon$ is the mechanical luminosity of the wind in units of $10^{36}$ erg s$^{-1}$, $n$ is the ISM density around the star (in cm$^{-3}$), $t$ is the age of the shell in units of $10^4$ yr and $c_{HII}$ is the sound velocity (in km s$^{-1}$). Eq. (4) assumes a mass per particle $\mu = 0.6m_p$ corresponding to a fully ionized gas with solar abundance. The numerical coefficient in eq.(5) assumes a temperature of $10^4$ K, and hence a sound speed $c_{HII} \sim 10 \ \mathrm{km\,s^{-1}}$ and an $H\alpha$ recombination coefficient of $8.7 \times 10^{-14} \ \mathrm{cm^{-3} \ s^{-1}}$ (Osterbrock 1989). The column density through the shell is given by

$$\Sigma_s = n_s \Delta R_s = 1.7 \times 10^{20} \epsilon^{1/2} n^{1/2} \mathcal{M}^{-1/6} \left[ \mathcal{M}^{2/3} - (\mathcal{M}^2 - 1)^{1/3} \right] \ \mathrm{cm^{-2}} \qquad (6)$$

For $n = 100 \ \mathrm{cm^{-3}}$ and $\epsilon = 1$ the shell column density increases from $1.8 \times 10^{19}$ to $1.7 \times 10^{21} \ \mathrm{cm^{-2}}$ as the shell slows down from $v_s = 100$ to $v_s = 10 \ \mathrm{km\,s^{-1}}$. Assuming that only about $10^{-4} - 10^{-5}$ of the H atoms are in the level n = 2 this yields such low values of the $H\alpha$ optical depth as clearly warrant the optically thin approximation.

The line profile produced by a constant velocity ($v_s$), optically thin spherical shell is *flat-topped* between $-v_s$ and $v_s$ (e.g. Beals 1931, see Figure 1). Thermal motions of the gas smooth the profile on scales of $\approx c_{HII}$, but otherwise the emission should be constant across the line profile, with an intensity given by $F_s(v) = L_s/2v_s$ (see Figure 1).

A number of effects could in principle alter this profile shape. *(i)* At large optical depths radiative transfer effects come into play and alter the shape of the line profile. However, the $H\alpha$ optical depth is small, and so the optically thin approximation is a valid assumption. *(ii)* If the radial velocity is not constant across the shell then a different profile would result. Such velocity differences are not expected in wind-driven shells given their small geometrical thickness, and thus small velocity gradients would not be sufficient to produce significant profile changes. *(iii)* Non-spherical, partly broken or one-sided shells are often observed in giant HII regions. Sections of a spherical shell ("caps") still produce a flat top profile, but the velocity range would be smaller than $-v_s \to v_s$, and dependent on the orientation of the shell section. In a system containing many shells, however, one would expect a shell section oriented in one direction to be compensated by another shell section oriented in the opposite direction. The final effect would be that the profile resulting from a number of incomplete shells is equivalent to the profile of a single complete shell. This equivalence should apply unless the overall symmetry of the system is not spherical. *(iv)* Large shells in nearby giant HII regions are often spatially resolved, and their line profiles show line-splitting unless the whole shell is included in the spectrograph aperture.

## 3.2. Multiple shells

The line profile resulting from the superposition of many shell profiles can be computed using the formalism developed by Cid Fernandes & Terlevich (1994) and Cid Fernandes (1995). In a system containing many wind-driven shells ($N_s$), the total line profile will be the sum of many flat-top profiles, each with a different width and intensity, as these corresponds to different ages, mechanical luminosities and ISM densities:

$$F_{tot}(v) = \sum_{i=1}^{N_s} F_s(v; t_i, \epsilon_i, n_i) \qquad (7)$$

For a large number of shells, this summation can be approximated by an integral over the distribution of $t$, $\epsilon$ and $n$:

$$F_{tot}(v) \approx \int F_s(v; t, \epsilon, n) dN_s(t, \epsilon, n) = N_s \int \int \int F_s(v; t, \epsilon, n) p(t, \epsilon, n) dt \, d\epsilon \, dn \qquad (8)$$

where $N_s$ is the total number of shells and $p(t, \epsilon, n)$ denotes the probability density of a given set of $t$, $\epsilon$ and $n$. A large variety of profile shapes can be obtained by arbitrarily choosing the probability distribution function, and each of these (see below) results into a very precise shape of the total line profile.

### 3.2.1. Time-averaged line profiles

The probability $p(t, \epsilon, n)$ can be written as $p(t|\epsilon, n)p(\epsilon, n)$; *i.e.*, the product of the probability that a shell with a given $\epsilon$ and $n$ is present at a time $t$ and the distribution function of $\epsilon$ and $n$ among all shells. Eq. (8) then becomes

$$F_{tot}(v) = N_s \int \int F_{tot}(v; \epsilon, n)p(\epsilon, n)d\epsilon \, dn \qquad (9)$$

where $F_{tot}(v; \epsilon, n)$, is the average profile of shells with a given $\epsilon$ and $n$ but different ages, given by

$$F_{tot}(v; \epsilon, n) = \int_{t_i}^{\min(t_v, t_f)} F_s(v; t, \epsilon, n)p(t|\epsilon, n)dt \qquad (10)$$

where $t_i$ and $t_f$ are the initial and final ages of the shell respectively and $t_v$ is the age of a shell with velocity $v_s = v$ (older, slower shells do not contribute to the profile at $v$). A shell can only emit in H$\alpha$ if it is subjected to photoionizing radiation. Thus, a necessary requirement is that the shock causing the shell slows down to velocities $\approx 100\text{--}250 \text{ km s}^{-1}$ or less to ensure rapid recombination. Faster shocks (younger shells) lead to higher temperatures in the swept up matter, and given the interstellar cooling law enter a quasi-adiabatic regime that inhibits recombination. This sets the initial age to $t_i \approx t(v_s = 100 \text{ km s}^{-1}) = 10^4(\epsilon/n)^{1/2}$ yr. The final age $t_f$ should correspond to the time at which the shell velocity approaches the random speed of motions of the interstellar medium, losing its identity upon disruption (Spitzer 1968).

Regardless of the exact values of $t_i$ and $t_f$, replacing $F_s$ by $L_s/2v_s$ in equation 10 and using eqs. 1 and 5 yields the following solution for $F_{tot}(v; \epsilon, n)$:

$$F_{tot}(v; \epsilon, n) = \int_{t_i}^{\min(t_v, t_f)} \frac{L_s(t)}{2 v_s(t)} p(t|\epsilon, n) dt = A\epsilon^{4/5} n^{6/5} \int_{t_i}^{\min(t_v, t_f)} t^{7/5} p(t|\epsilon, n) dt \qquad (11)$$

where $A = 8.0 \times 10^{31} (c_{HII}/10)^{-2}$ erg s$^{-1}$/km s$^{-1}$. Therefore, the profile at velocity $v$ is proportional to the mean $t^{7/5}$ of the shells faster than $v$.

The exact distribution of ages is unpredictable without the detailed history of star formation in the system. Information on the time at which strong stellar winds begin to act for stars of different masses, and the span of time that they remain in full operation would also be required to evaluate $p(t|\epsilon, n)$. However, unless star formation is synchronized to less than the life-time of a shell ($\tau_s \approx 10^6$ yr for massive stars), it seems a reasonable approximation to assume all ages to be equally likely, with old shells from the first stages of star formation and young shells from newly born stars occurring in similar proportions. The probability of observing a shell with age between $t$ and $t + dt$ is therefore $p(t)dt = dt/\tau$, where $\tau = \tau_{sf} + \tau_s$ is the duration of the star-formation episode plus the lifetime of the shell. Both $\tau_{sf}$ and $\tau_s$ can in principle depend on $\epsilon$ and $n$. This distribution of ages yields the following solution to eq. (11)

$$F_{tot}(v; \epsilon, n) = \frac{5A\epsilon^{4/5} n^{6/5}}{12\tau} \left\{ [\min(t_v, t_f)]^{12/5} - t_i^{12/5} \right\} \qquad (12)$$

which, making use of the relation between $v_s$ and $t$ (eq. 1), is equivalent to

$$F_{tot}(v; \epsilon, n) = \frac{5A\epsilon}{12\tau} \left( \frac{v_i}{100} \right)^{-6} \left[ \left( \frac{\max(v, v_f)}{v_i} \right)^{-6} - 1 \right] \qquad (13)$$

Therefore, the total line profile expected from systems containing many wind driven shells with an uniform distribution of ages has a flat-top core between $-v_f$ and $v_f$, and decays as $v^{-6}$ up to the maximum detectable shell velocity, $v_i$ (see Figure 2). This very steep profile is a result of the rapid decay of $v_s$ with time and the even more rapid increase in line luminosity as $t$ increases, which combined give a much larger weight to older, slower shells. Note also that since *all* shells contribute to the line core and old shells contribute *only* to the line core, the total profile necessarily peaks at $v = v_f$.

The hypothesis that all ages between $t_i$ and $t_f$ occur with equal probability in multiple systems plays an important role in shaping $F_{tot}(v; \epsilon, n)$. Though this seems to be the most natural assumption, one might wonder what line profiles would result from different age distributions. In the extreme case where all the stars form simultaneously and all their winds act at the same time, all shells would be observed with the same age $t$, and $F_{tot}(v; \epsilon, n)$ would be identical to $F_s(v; t, \epsilon, n)$, i.e. a flat-top profile between $-v_s(t)$ and $v_s(t)$. A power-law distribution of ages with $p(t) \propto t^\beta$ would produce a profile similar to eq. (13), but with wings decaying as $v^{-6-5\beta/2}$ instead of $v^{-6}$.

One can also *invert* the problem and ask which distribution of ages/velocities would result from a chosen line profile. This is most easily done re-writing equation 11 as an integral over shell velocities instead of ages. Using eq. 1 one obtains

$$F_{tot}(v; \epsilon, n) = A 100^{7/2} \epsilon^{3/2} n^{1/2} \int_{\max(v_f, |v|)}^{v_i} v_s^{-7/2} p(v_s) dv_s \tag{14}$$

which can be inverted to

$$\frac{dF_{tot}(v_s)}{dv_s} = -A 100^{7/2} \epsilon^{3/2} n^{1/2} v_s^{-7/2} p(v_s) \tag{15}$$

This solution applies only for $|v| > v_f$, i.e. outside the line core. The profile between $-v_f$ and $v_f$ cannot be changed, being necessarily flat-topped irrespective of $p(v_s)$. The distribution of shell velocities required to produce a Gaussian profile with standard deviation $\sigma$ is shown in Figure. 3, along with the corresponding distributions of ages and the resulting line profile. (Note that the profile is Gaussian only outside the $-v_f$ to $+v_f$ core, where it remains flat-topped). In this case $p(v_s) \propto v_s^{9/2} \exp(-v_s^2/2\sigma^2)$, which peaks at $v_s = (9/2)^{1/2}\sigma = 2.12\sigma$. Thus, in order to obtain a Gaussian profile a very peaked distribution of shell ages is required. Such an age distribution implies a very contrived and hard to explain star-formation history. It could also imply a rather unlikely event given the short period of operation ($\leq 10^5$ yr) compared to the life-time of giant HII regions ($\leq 10^7$ yr).

In what follows we shall assume a uniform distribution of shell ages throughout the life-time of the nebulae, a possibility that seems a more natural distribution than the ones shown in Figure 3c.

### 3.2.2.   Variable $\epsilon$ and $n$

Strictly speaking, eq. (13) applies only to a population of shells with identical driving energies and evolving in a uniform-density medium. It is well known however, that wind-driven shells in giant HII regions have a variety of values of $\epsilon$ and clearly evolve into a highly non-uniform density medium. These could in principle modify the line profile shape. Note however that the total profile after convolving eq. (13) with $p(\epsilon, n)$ would change its $v^{-6}$ shape only if the initial and final velocities $v_i$ and $v_f$ have a dependance on $\epsilon$ or $n$. In fact, eq. (13) remains valid after $\epsilon/\tau$ is replaced by the mean value of this ratio among the various shells. If $p(\epsilon, n)$ is to affect the shape of the total line profile it has to do so by affecting $v_i$ or $v_f$, and thus one might as well replace $p(\epsilon, n)$ by $p(v_i, v_f)$. Given that the profile $F_{tot}(v; \epsilon, n)$ is so strongly concentrated around the terminal velocity, variations in $v_i$ from shell to shell would not have a strong impact on the total profile, and thus one can further simplify the problem by considering only variations in $v_f$. With these considerations the total line profile can be written as

$$
\begin{aligned}
F_{tot}(v) &= N_s \int_{v_f^{low}}^{v_f^{upp}} F_{tot}(v; v_f) p(v_f) dv_f \\
&= \frac{5AN_s}{12} \left(\frac{v_i}{100}\right)^{-6} \int_{v_f^{low}}^{v_f^{upp}} \left[\left(\frac{\max(v, v_f)}{v_i}\right)^{-6} - 1\right] \frac{\epsilon(v_f)}{\tau(v_f)} p(v_f) dv_f
\end{aligned} \tag{16}
$$

where $v_f^{low}$ and $v_f^{upp}$ define the range of possible $v_f$ values. To compute the total profile one needs to specify how $\epsilon$ and $\tau$ depend on $v_f$, besides the probability distribution of $v_f$ itself. Since these functions cannot be derived from first principles, a parameterized approach is required. Eq. 16 can be evaluated by defining a weight function $w(v_f) \equiv \epsilon(v_f)p(v_f)/\tau(v_f)$ and parameterizing $w$ as a power law: $w(v_f) \propto v_f^{\gamma}$. The resulting line profile is flat topped from $-v_f^{low}$ to $v_f^{low}$, and follows a $v^{-6}$ law for $|v| > v_f^{upp}$. For velocities between $v_f^{low}$ and $v_f^{upp}$ the profile is a combination of two power laws, one with a $-6$ slope and the other with a slope of $\gamma - 5$. Figure 4 illustrates this profile for different combinations of $\gamma$, $v_f^{low}$ and $v_f^{upp}$.

There are no strong conclusions to be derived from this analysis without a physical justification for the $w(v_f)$ expression. Nevertheless, Figure 4 already shows that unless the weight function is very strongly biased towards large terminal velocities, the total profile shape remains far from Gaussian. Roundish profiles begin to arise for $\gamma > 7$. Note, however, that even if one could physically justify such possible values of $\gamma$, a large dynamic range of $v_f$'s is also required to produce a total Gaussian-looking intrinsic profile.

### 3.3. Unresolved shells

For a given value of the ratio $\epsilon/n$, the constraint placed by spatial undetectability of wind-driven shells ($R_s \leq r_{crit}$) in giant HII regions can also be written as a limiting velocity criterion:

$$v_s \geq v_{crit} = 31 \left(\frac{\epsilon}{n}\right)^{1/3} \left(\frac{r_{crit}}{10 \text{ pc}}\right)^{-2/3} \text{ km/s}$$

Clearly, low values of $\epsilon/n$ (*i.e.* either low energy winds and/or high densities) favor the constraint as well as young, and thus small and/or fast moving, shells. The constraint is illustrated in Figure 5, where the minimum velocity ($v_{crit}$) such that shells are not spatially resolved are indicated for a range of $\epsilon/n$. Shells below the lines drawn for different $r_{crit}$ values would have exceeded $r_{crit}$ and thus be resolved. In the figure, the possible range of $\epsilon/n$ values to account for the 30 Doradus "slow shells" resolved by CK is clearly indicated for an assumed $r_{crit} = 10$pc. The range of values required to account for faster shells lies to the right and out of the plot, thus implying the joint action of many massive stars powering the detected fast expanding structures. Shells above the lines are unresolved. Note however that the rapid wind-driven shell evolution implies that unavoidably, most of the unresolved shells will soon have velocities close to $v_{crit}$. If one was to exclude, as demanded by the constraint, resolved shells from the computation of the total line profile arising from the background in giant HII regions, one could treat $v_{crit}$ as the terminal velocity $v_f$. This together with the values of velocity dispersion (15–40 km s$^{-1}$) and background density ($1 \leq n \leq 10^3$ cm$^{-3}$) found in giant HII regions implies rather low values of $\epsilon$, in fact indicative of the power expected from low mass stars ($E_{wind} \sim 10^{33-35}$erg s$^{-1}$) as shown by the shaded area in Figure 5. The above conclusion becomes more stringent the smaller the value of $r_{crit}$.

### 3.4. Synthetic observed profiles: thermal and instrumental effects

Here we consider the fact that observed line profiles are a convolution of intrinsic profiles with both the thermal motions of the photoionized gas and the instrumental profile. Both thermal and instrumental effects can be represented by Gaussians with a combined width given by $\sigma_G = (\sigma_{th}^2 + \sigma_{inst}^2)^{1/2}$, Typical values of $\sigma_{th}$ and $\sigma_{inst}$ are 10 km s$^{-1}$, which yield $\sigma_G = 14$ km s$^{-1}$. In principle, the intrinsic profile could be recovered by applying a deconvolution algorithm to the observed profile, but that is seldom done. Instead, the usual practice is to present the observed profiles and quote their corrected width $\sigma_{cor} = (\sigma_{obs}^2 - \sigma_G^2)^{1/2}$ under the assumption that all profiles considered behave as Gaussian distributions. Having this in mind, one can predict the shapes of observed profiles and their corrected width for different combinations of the intrinsic profile width and the expected values of $\sigma_G$.

Figure 6 shows a mosaic of profiles for different values of $v_f$ (which sets the width of the theoretical profile) and $\sigma_G$. The thin lines represent the theoretical intrinsic emission from a large number of unresolved shells, as calculated from equation 13. The dotted lines are the convolution of the thermal and instrumental profiles considered in each case, and the thick lines are the result of the convolution of the above mentioned lines, leading to the "synthetic observed profiles". The striking (albeit obvious) fact illustrated in this figure is that intrinsic profiles narrower than $v_f \approx 2\sigma_G$ are gaussianized by the thermal and instrumental effects. Clear departures from a Gaussian profile are only seen for $v_f \gtrsim 3$ $\sigma_G$. Note also that the values of $\sigma_{cor}$ are systematically smaller than the intrinsic width, defined by $v_f$. This is due to the convolution algorithm which by conserving flux leads, when considering a Gaussian and a steep profile (such as the $v^{-6}$ law), to a re-distribution of the flux towards the base of the profiles.

The figure allows for a direct assessment of the properties of the supersonic flow found in giant HII regions from the constraints that one can derive from the present study. Take for example the "quiescent flow" of 30 Doradus found as a Gaussian background component ($\sigma_{30Dor} \sim 18$ km s$^{-1}$) both in the core and the halo of the region (see CK), or the $\sigma_{604} \sim 17$ km s$^{-1}$ arising from the main emitting knots in NGC 604 (see Sabalisck *et al.* 1995). Given the expected thermal broadening in both regions ($\sim 10$ km s$^{-1}$) and the similar instrumental broadening ($\sigma_{inst} \sim 10$ km s$^{-1}$) in both studies one can derive a $\sigma_G \sim 14$ km s$^{-1}$. Thus the supersonic flow in the background of both regions can be well explained with the results shown in figure 6, in the panel with coordinates ($\sigma_G = 10$ km s$^{-1}$, $v_f = 20$ km s$^{-1}$). However, such a successful match immediately leads one to wonder why the terminal velocity ($v_f$) of the unresolved shells is supersonic in such regions, a fact that implies a supersonic random speed of motions of their local interstellar medium. Note that if $v_f$ was smaller, say 10 km s$^{-1}$, the resultant profile would be narrower, and therefore different to those observed in our nearest giant HII regions. The situation becomes even more critical if one considers the full range of $\sigma$ values found in giant HII regions, which extends up to 40 km s$^{-1}$. These, if produced in each case by a collection of unresolved wind-driven shells, takes us in our Figure 6 to even larger values of $v_f$ and thus to faster supersonic random speed of motions of the local ISM in these sources to account for the missing slower shells. Furthermore, if the thermal and instrumental broadening remain $\sim 10$ km s$^{-1}$, larger $v_f$'s also lead to wider flat-topped synthetic line profiles, in clear disagreement with the observations of giant HII regions which show a clear Gaussian behavior of the integrated emission lines across the whole range of supersonic $\sigma$ values (see Arsenault & Roy 1988). To gaussianize those line profiles it would be necessary to move towards the upper right-hand corner of Figure 6. This however, can only be achieved by assuming large changes in the thermal broadening, not only caused by a possible larger HII region temperature but rather due to the intrinsic supersonic random motions in each of the giant HII region sources. This brings us, once again, to infer the existence of another mechanism, different from the winds from massive stars, necessary to stir the ISM locally and thus explain the early disruption of unresolved wind-driven shells caused by massive stars while leading to a Gaussian integrated profile.

## 4. Results and discussion

Given the large number of massive stars causing the photoionization of giant HII regions and their large mechanical energy output, through strong stellar winds and supernova explosions, the volume of a nebula mechanically unperturbed should become largely reduced as a function of time, and thus the observed line profiles should evolve to be dominated eventually by the emission from matter affected by the disruptive action from massive stars. Under these conditions the unperturbed gas may be rescued only by means of high spatial and spectral resolution, and in evolved HII regions high sensitivity instrumentation would be particularly required. All objects subjected to such studies (30 Doradus, NGC 604, NGC 5471, NGC 5461, etc; see Muñoz-Tuñón 1994) show Gaussian profiles with a width, a $\sigma$ value, that implies a supersonic velocity dispersion in regions where the presence of strong stellar winds is not evident. We have explored the possibility that unresolved wind-driven shells produced by massive stars could account for the range of supersonic $\sigma$ values found in giant HII regions (15–40 $km\,s^{-1}$). The main results from our analysis are :

1) A collection of spatially unresolved, concurrently evolving wind-driven shells immersed in an HII region cannot cause a supersonic Gaussian line profile unless the wind-driven sources have a strongly peaked and thus peculiar age distribution (see Figure 3). This can also be interpreted as a short-live event and thus, given the presence of supersonic line widths in all giant HII regions, makes this possibility rather unlikely.

2) A uniform age distribution of the wind-driven sources leads to distinct non-Gaussian intrinsic line profiles and thus do not match the observations. These present a flat-topped core defined by the range of terminal shell velocities (*i.e.* from $-v_f$ to $v_f$), and a steep decay as $v^{-6}$ up to the largest detectable shell speed ($v_i$).

3) Spatial undetectability of wind-driven shells is clearly favored by large background densities and low energy winds. For the closest giant HII regions, such as 30 Doradus where the supersonic gas is found both in the core of the region (with typical densities $\sim 10^3\ cm^{-3}$) and in the halo (where the typical density is $\sim 1$ cm$^{-3}$), to have a collection of unresolved shells causing the supersonic width of the observed lines implies rather low-energy winds with properties similar to those typical of low-mass stars ($E_{wind} \sim 10^{33-35}$ erg $s^{-1}$).

4) At the same time, the conditions that produce an intrinsic supersonic Gaussian profile while accounting for the unresolved shells as well as for a large range of values of wind energetics and background densities ($\epsilon/n$), lead unavoidably to large final shell velocities ($v_f$). These imply supersonic values of the random speed of motions of the local interstellar medium disrupting the evolving shells long before they are capable of slowing down to speeds below 10 km s$^{-1}$. To our knowledge, the only possibility of causing and maintaining a supersonic value of the random speed of motions is through the continuous passage of a collection of low-mass wind-driven stars and their cometary shocks ( Tenorio-Tagle *et al.* 1993). In such a case the motion of the ensemble is directly related to the total mass of the system, while the energy causing the stirring comes from the winds of the low-mass stars.

5) Thermal and instrumental broadening tend to gaussianize line profiles, and the effect is more noticeable the narrower the intrinsic profiles are with respect to the Gaussian filter ($\sigma_G$). Note however, that the convolution of a narrow line arising from a system of unresolved shells and a Gaussian filter may lead to a broad Gaussian profile with a supersonic $\sigma$ value only if the final speed of shells (that defines the width of the flat-topped core of the intrinsic profile) is also supersonic. This implies that the local random speed of motions of the target giant HII region is also supersonic, and thus capable of disrupting the evolving shells when $v_s = v_f$ long before they reach, as in the Galaxy, a speed of $\leq 10$ km s$^{-1}$. Broader intrinsic lines (with $v_f \geq 30$ km s$^{-1}$) would however appear as described in 2), and the only way to make such profiles acquire a Gaussian shape is through a convolution with a broader filter. The expected thermal and instrumental widths are, however, well determined and thus, the only remaining possibility is to add to the Gaussian filter a supersonic component representing the local random speed of motions of the observed HII region.

Thus our attempt to explain the supersonic Gaussian lines observed in giant HII regions with a large collection of unresolved wind-driven shells from massive stars leads in all cases to the existence of an underlying supersonic turbulent motion, independent of the wind-driven shells caused by massive stars. Furthermore, this underlying supersonic motion is what limits and defines the different terminal speed of shells in each of the giant HII regions. In our approach we had assumed static wind driven sources developing the unresolved shells, although clearly an alternative way to gaussianize a broad emission with an intrinsic $v^{-6}$ behavior is to allow for the motion of the sources within the gravitational potential of the whole (stars + gas) system. However, given the energetics of stellar winds and the densities of giant HII regions, only in the case of low-mass stars will the stellar wind configurations be transformed into an ensemble of bow, or cometary, shocks raming through the nebula. These, while stirring the photoionized gas causing a supersonic random speed of motions, lead to the observed broad Gaussian line profiles.

A further indication that winds from low-mass stars are acting over large volumes comes from the often disregarded fact that the HI clouds associated with giant HII regions also show a comparable supersonic width, and both 30 Doradus and NGC 604 are good examples of this. McGee & Milton (1966) determined a $\sigma$ value of 15.8 km s$^{-1}$ for their position (L32) which encompasses the massive ($1.4 \times 10^7$ M$_\odot$), 500pc diameter HI cloud associated to 30 Doradus. There are also several other regions (positions L43, 48, 49, 50) with a supersonic velocity dispersion all located in the neighboring kpc of 30 Doradus. Position L32 is however the most intense of these lines by more than a factor of 10. In the case of NGC 604, Wright (1971) measured a $\sigma$ of 23 km s$^{-1}$, also largely supersonic for an HI cloud. There are many other examples. Viallefond *et al.* (1981) determined and compared the HI and HII supersonic $\sigma$ values of the giant regions in M101 (NGC 5471, NGC 5462, NGC 5461), Hodge 58, NGC 5455, and NGC 5447, and all turn up to be rather similar.

The measured $\sigma_{HI}$ implies a large amount of kinetic energy in the HI phase, energy that cannot be attributed to the energetics from massive stars. The close agreement between the HI $\sigma$ values and those derived for the associated HII region gas, indicate therefore a different source of energy.

All of these facts, *i.e.* the favoured low values of $\epsilon$, the large final shell velocities or the implied supersonic random speed of motions of the local ISM, and the similar values of $\sigma$ found for the ionized and neutral components of each giant HII region, are elements contemplated in the cometary stirring model proposed by Tenorio-Tagle , Muñoz-Tuñón & Cox (1993). The model predicts $\sigma_{gas} \sim \sigma_{stars}$, as has been confirmed for the Orion cluster and the Orion molecular cloud (see review of Genzel & Stutzki 1989). For giant HII regions however, this fact still awaits ratification.

We thank our referee for many suggestions to improve the presentation, and to Drs. E. Perez, R. Terlevich and Prof. G. Munch for his comments and suggestions. Our thanks also to Dr. T Mahoney for careful reading of the manuscript. We acknowledge partial financial support from the EEC through the ANTARES grant for collaborative research. GTT and CMT also acknowledge partial support from DGICYT (grant PB91-0531GEFE). RCF acknowledges the Brazilian agency CAPES for financial support through grant 417/90-8.

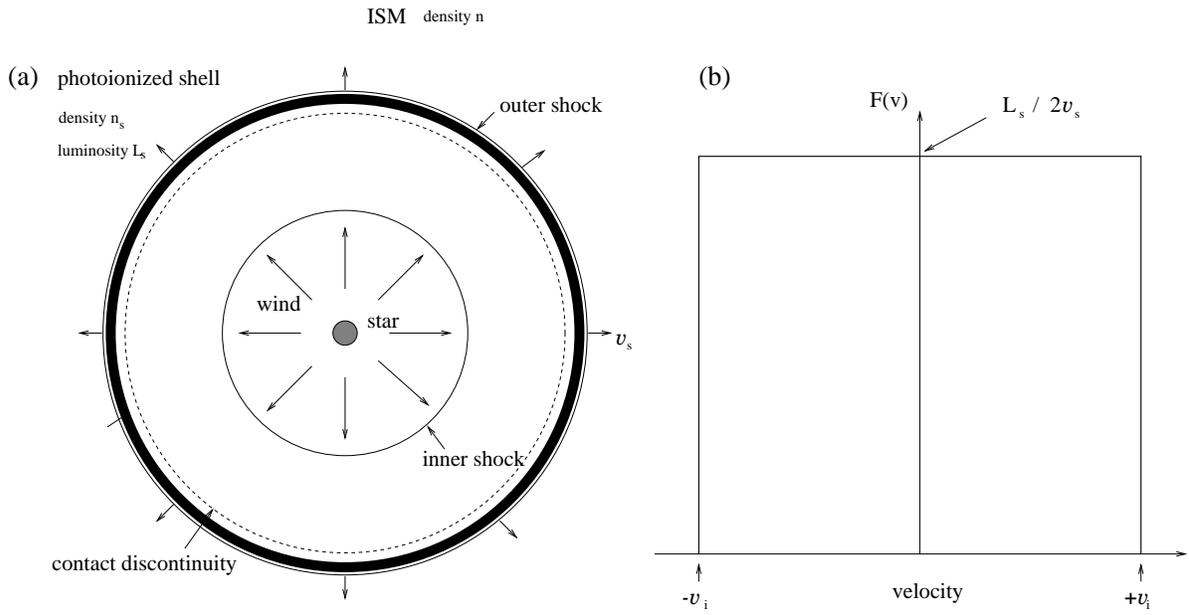

Fig. 1.— Schematic representation of a wind-driven shell and its flat top emission profile.

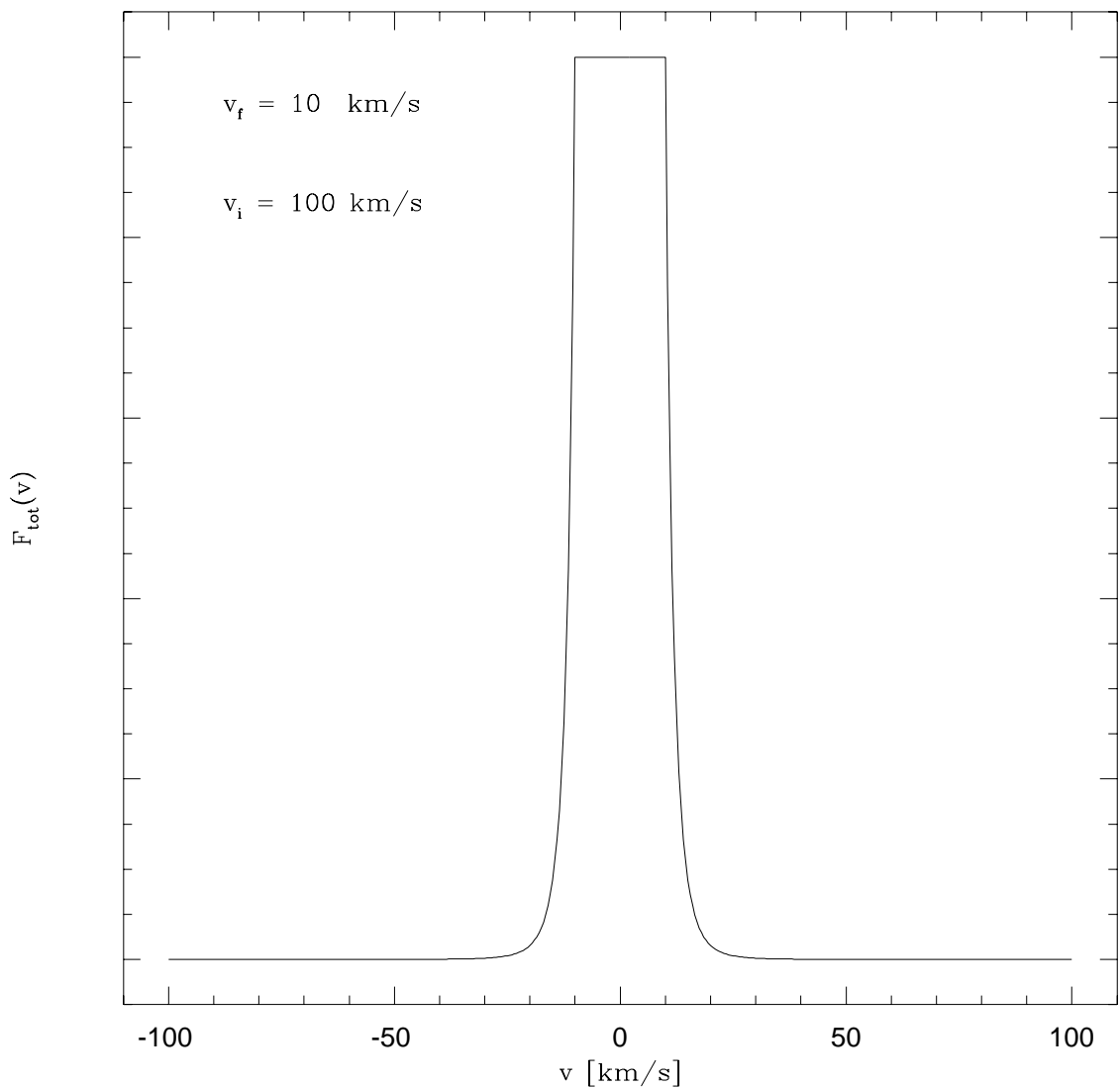

Fig. 2.— Unresolved wind-driven shells. The emission line profile resulting from a collection of photoionized unresolved wind-driven shells with a uniform age distribution and initial and final speeds of 100 and 10 km s$^{-1}$, respectively. Note that larger values of the initial speed would hardly affect the resultant profile.

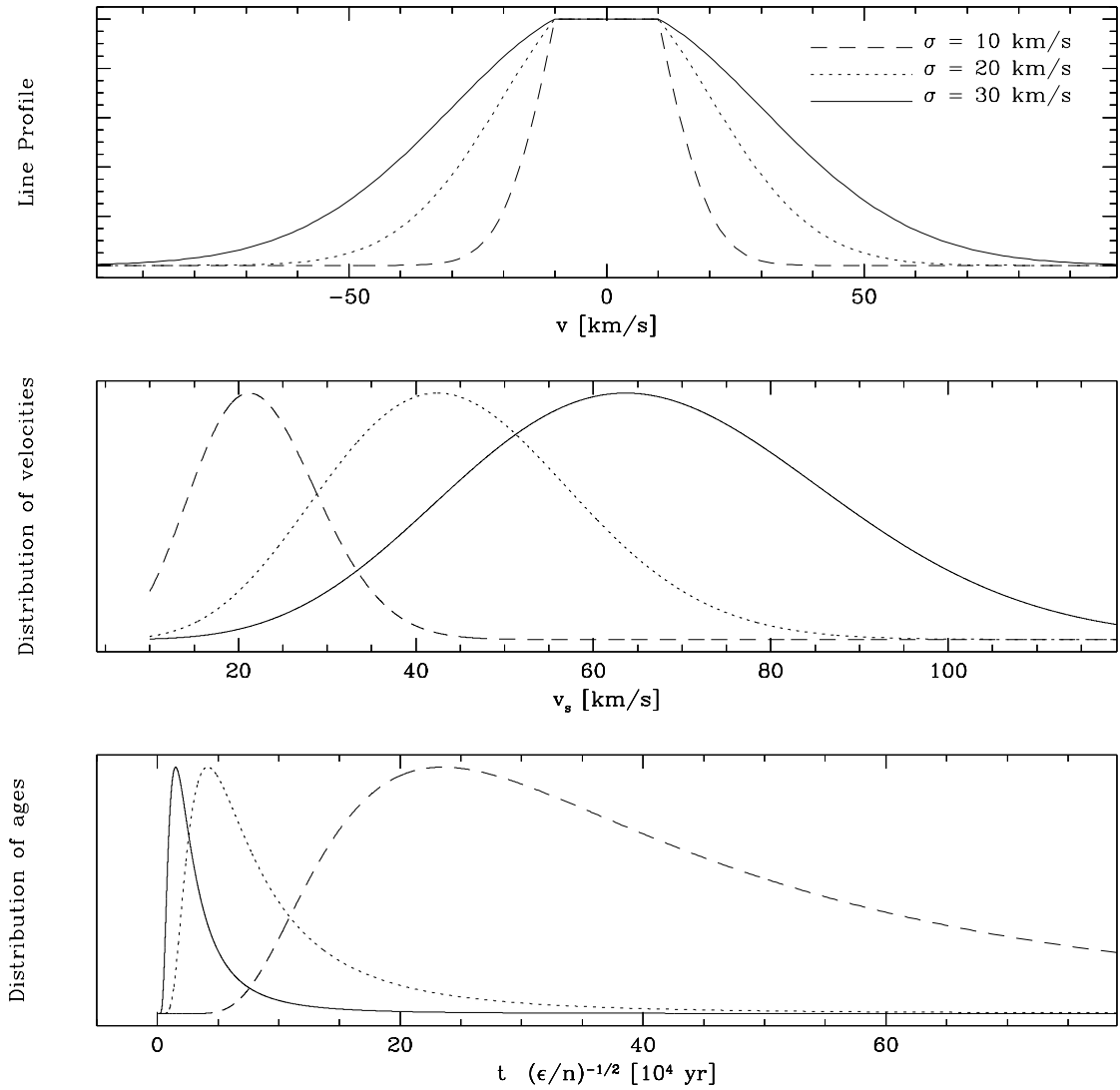

Fig. 3.— Unresolved wind-driven shells. Distribution functions for shell ages (top) and corresponding velocities (center) such that the total intrinsic line profile is a Gaussian with standard deviation $\sigma = 10$, 20 and 30 $km\,s^{-1}$(lower panel).

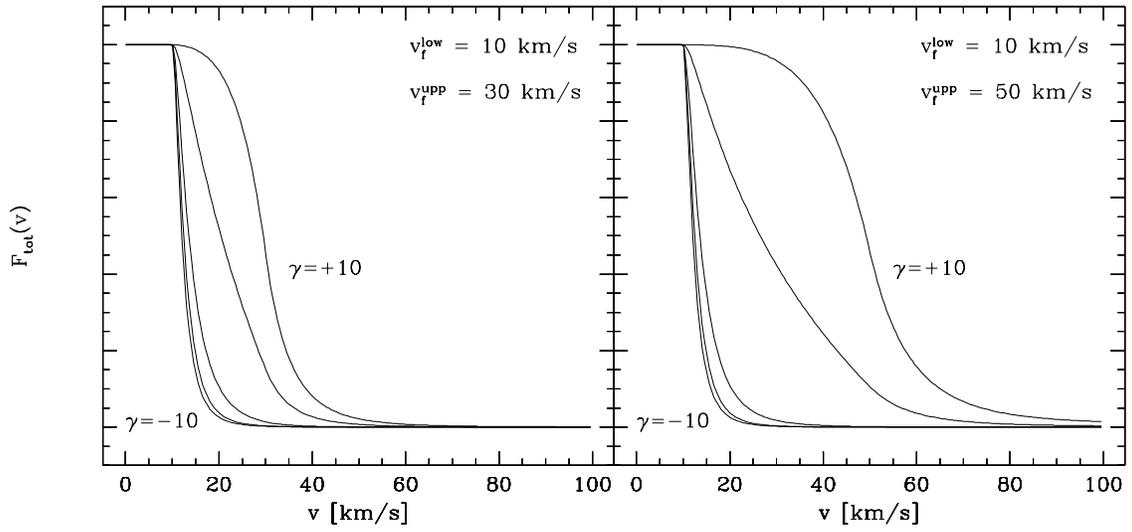

Fig. 4.— Wind-driven shells. Total intrinsic line profiles resulting for a system of shells with different terminal velocities between $v_f^{low}$ (10 km s$^{-1}$) and $v_f^{upp}$ (30 and 50 km s$^{-1}$), and a power-law weight function.

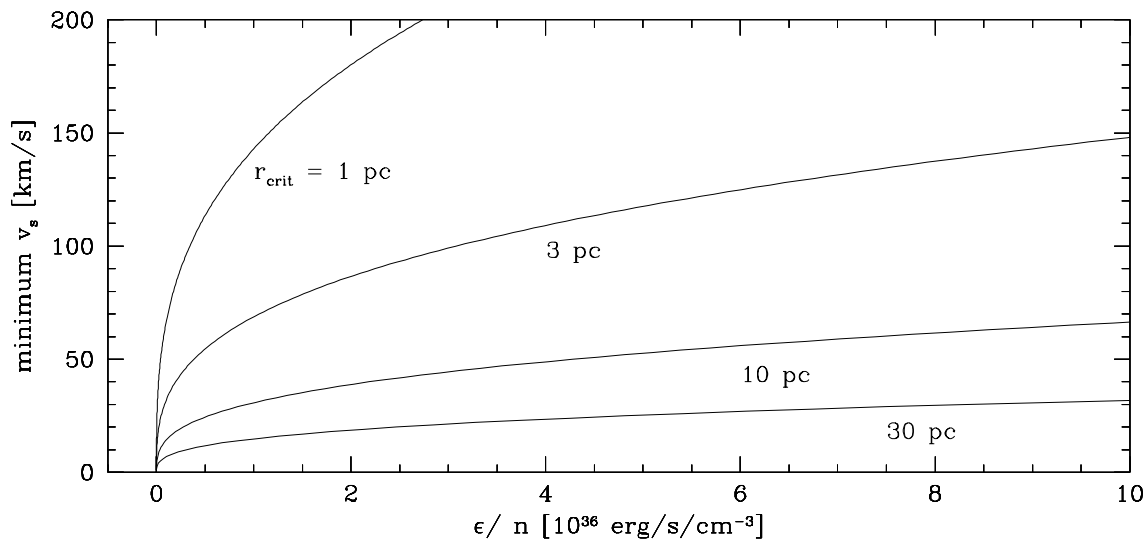

Fig. 5.— Wind-driven shells. Minimum shell velocity ($v_s$) plotted against the ratio of $\epsilon/n$ (in units of $10^{36}$ erg $s^{-1}$). Shells become detectable as they exceed the $r_{crit}$ value determined by the solid lines and cross into the area below the curves. The range of $\epsilon/n$ (assuming an $r_{crit}$ = 10 pc) for the detected slow expanding shells detected in 30 Doradus ($30 \leq 50$ km s$^{-1}$) is indicated by the dashed area, while unresolved shells covering the range of $\sigma$ values detected in giant HII regions (15–40 km s$^{-1}$) is indicated by the dotted area. Note that the latter becomes smaller, causing extreme values of $\epsilon/n$ as the spatial resolution increases (*i.e.* for smaller values of $r_{crit}$)

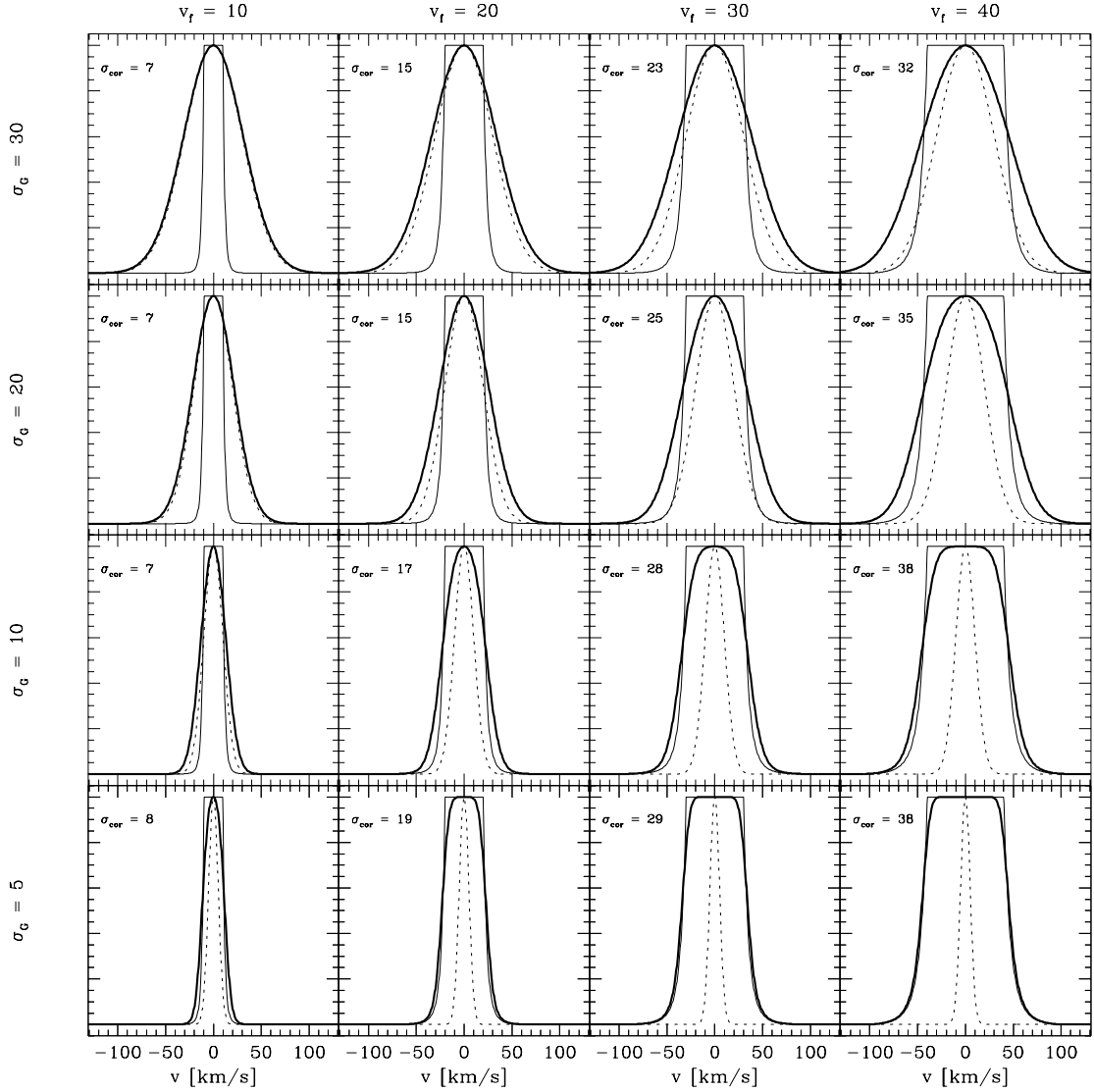

Fig. 6.— Synthetic line profiles. Predicted $v^{-6}$ line profiles (narrow lines) for a multiple-shell system convolved with a thermal plus instrumental Gaussian (dotted lines) of different widths. The thick lines show the resulting normalized line profile after convolution. The corrected values of $\sigma$ are indicated in each frame.